\documentclass[prl,twocolumn,amsmath,amssymb,superscriptaddress]{revtex4-2}

\usepackage[latin9]{inputenc}
\usepackage{epsfig}
\usepackage[T1]{fontenc} 

\usepackage{url}
\usepackage{ucs}
\usepackage{bbm}

\usepackage{hyperref}
\hypersetup{colorlinks=true, citecolor=blue, urlcolor=blue, linkcolor=blue}

\newcommand{\bra}[1]{\langle #1|}
\newcommand{\ket}[1]{|#1\rangle}
\newcommand{\ketbra}[1]{| #1\rangle \langle #1|}

\newcommand{\be}{\begin{equation}}
\newcommand{\ee}{\end{equation}}
\newcommand{\eea}{\end{eqnarray}}
\newcommand{\bea}{\begin{eqnarray}}

\newcommand{\va}[1]{\ensuremath{(\Delta#1)^2}}

\newcommand{\ex}[1]{\ensuremath{\langle{#1}\rangle}}
\newcommand{\exs}[1]{\ensuremath{\langle{#1}\rangle}}

\newcommand{\eins}{\mathbbm{1}}
\newcommand{\qed}{\ensuremath{\hfill \blacksquare}}

\newcommand{\kommentar}[1]{}
\newcommand{\trace}{{\rm Tr}}

\newcommand{\forget}[1]{}

\newcommand{\EQ}[1]{Eq.~\eqref{#1}}
\newcommand{\EQS}[1]{Eqs.~\eqref{#1}}
\newcommand{\EQL}[1]{Equation~\eqref{#1}}

\newcommand{\FIG}[1]{Fig.~\ref{#1}}
\newcommand{\REF}[1]{Ref.~\cite{#1}}
\newcommand{\REFS}[1]{Refs.~\cite{#1}}

\newcommand{\CITESUPP}{\cite{Note1}}

\newcounter{myequation}
\makeatletter
\@addtoreset{equation}{myequation}
\makeatother

\newcounter{myfigure}
\makeatletter
\@addtoreset{figure}{myfigure}
\makeatother

\newcounter{mytable}
\makeatletter
\@addtoreset{table}{mytable}
\makeatother

\begin{document}

\title{Quantum states with a positive partial transpose are useful for metrology}

\author{G\'eza T\'oth}
\email{toth@alumni.nd.edu}
\homepage{http://www.gtoth.eu}
\affiliation{Department of Theoretical Physics,
University of the Basque Country
UPV/EHU, P.O. Box 644, E-48080 Bilbao, Spain}
\affiliation{IKERBASQUE, Basque Foundation for Science,
E-48013 Bilbao, Spain}
\affiliation{Wigner Research Centre for Physics, Hungarian Academy of Sciences, P.O. Box 49, H-1525 Budapest, Hungary}
\author{Tam\'as V\'ertesi}
\email{tvertesi@atomki.hu}
\affiliation{Institute for Nuclear Research, Hungarian Academy of Sciences, P.O. Box 51, H-4001 Debrecen,  Hungary}

\begin{abstract}
We show that multipartite quantum states that have a positive partial transpose with respect to all bipartitions of the particles can outperform separable states in linear interferometers. We introduce a powerful iterative method to find such states. We present some examples for multipartite states and examine the scaling of the precision with the particle number. Some bipartite examples are also shown that possess an entanglement very robust to noise. We also discuss the relation of metrological usefulness to Bell inequality violation. We find that quantum states  that do not violate any Bell inequality can outperform separable states metrologically. We present such states with a positive partial transpose, as well as with a non-positive positive partial transpose.

\vspace{1em}
\noindent DOI: \href{https://doi.org/10.1103/PhysRevLett.120.020506}{10.1103/PhysRevLett.120.020506}
\end{abstract}

\date{\today}

\maketitle

\section{Introduction}

Entanglement lies at the heart of quantum mechanics and plays an important role in quantum information theory \cite{Horodecki2009Quantum,*Guhne2009Entanglement}.
However, in spite of intensive research, many of the intriguing properties of entanglement are not fully understood. One of such puzzling facts is that, while entanglement is a sought after resource, not all entangled states are useful for some particular quantum information processing application. For instance, in the Ekert protocol for quantum cryptography \cite{Ekert1991Quantum}, entangled states that do not violate a Bell inequality are not useful. Moreover, maximally entangled singlets cannot be distilled from entangled quantum states that have a positive-semidefinite partial transpose
(PPT). Such states, called also bound entangled, have been at the center of attention in quantum information science \cite{Peres1996Separability,Horodecki1997Separability}.

Recently, it has been realized that entangled states can be useful in very general metrological tasks in the sense that they make it possible to overcome the shot-noise limit in the precision of parameter estimation corresponding to classical interferometers \cite{Pezze2009Entanglement,Hyllus2012Fisher,*Toth2012Multipartite,Lucke2011Twin,Krischek2011Useful,Strobel2014Fisher}. Notably, separable states, i.e., states without entanglement cannot overcome the classical limit. However, again, there are highly entangled states that are not useful for metrology \cite{Hyllus2010Not}.

The relation between the various subsets of entangled states have been studied for a long time. It has been conjectured by A. Peres that no bound entangled state violates a Bell inequality \cite{Peres1999All}, which, after numerous attempts, has been recently refuted \cite{Vertesi2014Disproving}. The search for counterexamples has been hindered by the fact that the conjecture is very close to be true. At this point the questions arise: are there bound entangled states that are metrologically useful? Can states that do not violate any Bell inequality be metrologically useful? Finding such states numerically seems to be as easy as finding a needle in the haystack, since we need to {\it maximize} a convex function over a convex set.  There have been results concerning entanglement criteria with several quantum Fisher information terms detecting PPT entangled states as well as concerning the metrological usefulness of multipartite states that are not PPT with respect to all bipartions \cite{Hyllus2012Fisher,Altenburg2017PhDThesis, Czekaj2015Quantum}. However, it is a famously hard open problem of quantum information theory whether states with only PPT entanglement can be useful for metrology \cite{Czekaj2015Quantum}.

\begin{figure}[t!]
\centerline{
\epsfxsize5.5cm \epsffile{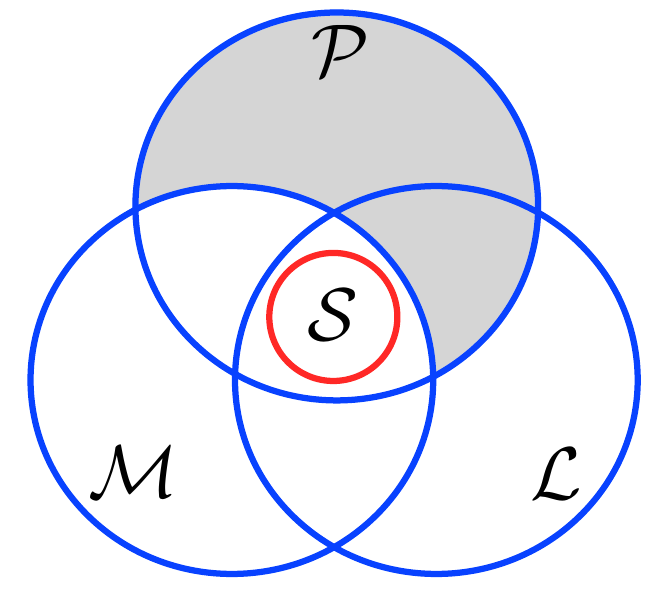}
}
\caption{Various convex sets of quantum states represented by circles: ($\mathcal{P})$ PPT states, ($\mathcal{M}$)  states that are not useful for metrology, ($\mathcal{S}$) separable states, ($\mathcal{L}$) states with a local hidden variable model. (grey area) Metrologically useful PPT states.  Such states are in $\mathcal{P}\backslash\mathcal{M},$ 
where ``$\backslash$'' denotes the difference between two sets.
} \label{fig:sets}
\end{figure}

In this paper, we give an affirmative answer to the question above. We show that there are bound entangled states that outperform all separable states metrologically, as depicted in \FIG{fig:sets}. Below, we summarize the four main contributions of this paper. 

(i) We present multiqubit quantum states that are metrologically useful, while having a positive partial transpose with respect to all bipartitions. In this way, we make sure that the metrological advantage compared to separable states cannot be attributed to the non-PPT bipartitions. 

(ii) We also present several bipartite PPT states for dimensions from $3\times3$ to $12\times12$ that  outperform separable states in quantum metrology. The metrological advantage of these states compared to separable states is very robust to noise. Thus, such states might be realized in experiments with photons or trapped cold ions (e.g., \cite{Kiesel2005Experimental,Walther2005ExperimentalOneWay,Haffner2005Scalable,Leibfried2005Creation}). 

(iii) We show an iterative method based on semidefinite programming (SDP) that can generate such states very efficiently. The method, starting from a given initial state, provides a series of PPT quantum states with a rapidly increasing metrological usefulness. 

(iv) We now turn to the relation of metrological usefulness and other convex sets of quantum states. We show that quantum states with a local hidden variable model, i.e., not violating any Bell inequality can be metrologically useful. We present such states with a positive as well as with a non-positive partial transpose \cite{Czekaj2015Quantum}.  

{\it Quantum Fisher information.}---Before discussing our main results, we review some of the fundamental relations of quantum metrology. A basic metrological task in a linear interferometer is  estimating the small angle $\theta$ for a unitary dynamics $U_{\theta}=\exp(-iA\theta),$ where $A=\sum_{n=1}^N a^{(n)},$ $N$ is the particle number and $a^{(n)}$ are single particle operators. The precision is limited by the Cram\'er-Rao bound as \cite{Helstrom1976Quantum,*Holevo1982Probabilistic,*Braunstein1994Statistical,
*Petz2008Quantum,*Braunstein1996Generalized,Giovannetti2004Quantum-Enhanced,*Demkowicz-Dobrzanski2014Quantum,*Pezze2014Quantum,*Toth2014Quantum,Paris2009QUANTUM}
\be \label{eq:cramerrao}
\va{\theta}\ge1/\mathcal{F}_{\rm Q}[\varrho,A],
\ee
where the quantum Fisher information, a central quantity in quantum metrology is defined by the formula \cite{Helstrom1976Quantum,*Holevo1982Probabilistic,*Braunstein1994Statistical,
*Petz2008Quantum,*Braunstein1996Generalized}
\begin{equation}
\label{eq:FQ}
F_{Q}[\varrho,A]=2\sum_{k,l}\frac{(\lambda_{k}-\lambda_{l})^{2}}{\lambda_{k}+\lambda_{l}}\vert \langle k \vert A \vert l \rangle \vert^{2}.
\end{equation}
Here, $\lambda_k$ and $\ket{k}$ are the eigenvalues and eigenvectors, respectively, of the density matrix $\varrho,$ which is used as a probe state for estimating $\theta$ .

It has been shown that for separable multi-qubit states the quantum Fisher information, characterizing the maximal precision achievable by a quantum state is bounded as \cite{Pezze2009Entanglement}
\be
\mathcal{F}_{\rm Q}[\varrho,J_z]\le N,\label{eq:Fsep}
\ee
where $J_z=\sum_{n=1}^{N} j_z^{(n)},$ and $j_z^{(n)}$ are the single particle angular momentum components.   \EQL{eq:Fsep} can easily be generalized for qudits with a dimension $d>2$ and operators $A$ different from $J_z$. If $\mathcal{F}_{\rm Q}[\varrho,A]$ is larger than the maximum reached by separable states then  $\varrho$ is useful for metrology. The maximum for separable states is given by $\sum_{n=1}^N [\lambda_{\max}(a^{(n)})-\lambda_{\min}(a^{(n)})]^2,$ where $\lambda_{\min}(a^{(n)})$ and $\lambda_{\max}(a^{(n)})$ denote the minimum and maximum eigenvalues of $a^{(n)},$ respectively \CITESUPP.

{\it Main results}---Is \EQ{eq:Fsep} also valid for PPT states, i.e., multiqubit states that are PPT with respect to all partitions? One could expect that this is the case since PPT states can only be weakly entangled, while they are highly mixed. The latter property hinders the violation of \EQ{eq:Fsep}  since the  Fisher information is convex, decreasing strongly after mixing quantum states.

Next, we present our first main result. We show that states with a positive partial transpose can still violate \EQ{eq:Fsep} and its generalizations for $A\ne J_z.$ Now we give concrete examples, mentioning first only the main properties of the states found numerically.

{\it Four-qubit state.---}PPT with respect to all bipartitions, and
with $A=J_z.$ 

{\it Three-qubit state.---}PPT with respect to all bipartitions.
We consider operators different from $J_z,$ and take $A=j_z^{(1)}+j_z^{(2)}.$

{\it Qubit-ququart bipartite PPT entangled state.---}It is a three-qubit state for which only the $1:23$ partition is PPT, while the other two bipartions are not PPT. Hence, the state has a higher value of quantum Fisher information than the three-qubit PPT state presented before. The three-qubit state can easily be transformed into a  $2\times 4$ bound entangled state, having the smallest dimensions in which PPT entanglement is possible.  

{\it Bipartite states of two qudits with equal dimension.---}$d\times d$ states with $d=3,4,\ldots,12,$ with the operator
\be
A=\openone \otimes D + D \otimes \openone, \quad D={\rm diag}(1,1, \ldots,-1,-1),\label{eq:Doperator}
\ee
where for even $d$ in the diagonal of $D$ there are $d/2$ $1$'s and $d/2$ $-1$'s, and for odd $d$ there are $(d+1)/2$ $1$'s and $(d-1)/2$ $-1$'s.

The quantum Fisher information of the states found, with other relevant properties, are summarized in Tables~\ref{tab:FQresults} and \ref{tab:FQresults_bipartite}. The density matrices of all the states are available in the Supplemental Material \footnote{See Supplemental Material for additional results about maximizing the negativity and computing the robustness, as well as for the states found numerically.  The Supplemental Material  includes \REFS{DiVincenzo2003Unextendible,Breuer2006Optimal,Toth2015Evaluating,Sentis2016Quantitative,pptmetro_note,Toth2013Extremal,Wigner1963INFORMATION,Raussendorf2001A,Mandel2003Controlled,Scarani2005Nonlocality,Guhne2005Bell,Toth2006Two-setting}.}. 

\nocite{DiVincenzo2003Unextendible,Breuer2006Optimal,Toth2015Evaluating,Sentis2016Quantitative,pptmetro_note,Toth2013Extremal,Wigner1963INFORMATION,Raussendorf2001A,Mandel2003Controlled,Scarani2005Nonlocality,Guhne2005Bell,Toth2006Two-setting}

\begin{table}[hthb]
\begin{center}
\begin{tabular}{|c|c|l|l|l|}
\hline
System & $A$ & $ \mathcal{F}_{\rm Q}[\varrho,A]$ & $\mathcal{F}_{\rm Q}^{({\rm sep})}$ & $p_{\rm white\,noise }$\\
\hline
\hline
four qubits & $J_z$ & $4.0088$& $4$ & $0.0011$\\
\hline
three qubits & $j_z^{(1)}+j_z^{(2)}$ & $2.0021$& $2$& $0.0005$\\
\hline
\begin{tabular}{c} $2\times4$ \\ (three qubits, \\  only $1:23$ is PPT) \end{tabular}
& $j_z^{(1)}+j_z^{(2)}$ & $2.0033$& $2$ & $0.0008$ \\ \hline
\end{tabular}
\end{center}
\caption{Quantum Fisher information for PPT states found numerically in various systems. For each system, the maximum for separable states is shown. The robustness of the metrological usefulness of the states is also given, assuming white noise.
}
\label{tab:FQresults}
\end{table}

\begin{table}[hthb]
\begin{center}
\begin{tabular}{|l|l|c|l|}
\hline
$d$ & $ \mathcal{F}_{\rm Q}[\varrho,A]$ & $p_{\rm white\,noise}$ & $p^{\rm LB}_{\rm noise}$
\\
\hline
\hline
$3$ & $8.0085$ & $0.0006$& $0.0003$\\
\hline
$4$ & $9.3726$ & $0.0817$& $0.0382$\\
\hline
$5$ & $9.3764$ & $0.0960$& $0.0361$\\
\hline
$6$ & $10.1436$ & $0.1236$& $0.0560$\\
\hline
$7$ & $10.1455$ & $0.1377$& $0.0086$\\
\hline
$8$ & $10.6667$ & $0.1504$& $0.0670$\\
\hline
$9$ & $10.6675$ & $0.1631$& $0.0367$\\
\hline
$10$ & $11.0557$ & $0.1695$& $0.0747$\\
\hline
$11$ & $11.0563$ & $0.1807$& $0.0065$\\
\hline
$12$ & $11.3616$ & $0.1840$& $0.0808$\\
\hline
\end{tabular}
\end{center}
\caption{Quantum Fisher information for PPT states found numerically in two-qudit systems of local dimension $d,$ where the maximum of the quantum Fisher information for separable states is $8$. The operator $A$ is given in \EQ{eq:Doperator}. For each state, the robustness of the metrological usefulness of the states is shown for white noise. A lower bound on the tolerated separable noise is also given. 
}
\label{tab:FQresults_bipartite}
\end{table}

{\it Maximization over PPT states.}---We now describe the method that has been used to find the metrologically useful PPT states. Brute force maximization of the quantum Fisher information \eqref{eq:FQ}  for PPT states is extremely difficult, since it is a convex function of the state. Hence, the maximum will be taken on the boundary of the set of PPT states, and no method can guarantee to find the global optimum.

We look for a simpler solution. We know that the error propagation formula gives the precision of estimating the parameter $\theta$ by measuring the expectation value of the operator $M$ as
\be
\va{\theta}=\frac{\va{M}}{\vert \partial_\theta \ex{M}\vert^2}=\frac{\va{M}}{\ex{i[M,A]}^2_\varrho}.
\label{eq:vartheta}
\ee
Based on \EQ{eq:cramerrao}, the quantity $1/\va{\theta}$ provides a lower bound on the quantum Fisher information. Note that for some $M$ the bound is saturated \cite{Paris2009QUANTUM,Frowis2015Tighter,Apellaniz2015Detecting}.

We now show that it is possible to obtain the quantum state minimizing \EQ{eq:vartheta}.

{\bf Obvservation 1.} The minimum of the precision \eqref{eq:vartheta} for PPT states for a given operator $M$ can be obtained by a semidefinite program.

{\it Proof.}---Let us define first
\begin{eqnarray}
f_M(X,Y)=\min_{
\varrho}&&
 {\rm Tr}(M^2\varrho),\nonumber\\
\textrm{s.t. }&&
\varrho\ge 0,\varrho^{{\rm T}k}\ge 0 \text{ for all }k, {\rm Tr}(\varrho)=1, \nonumber \\
&&  \ex{i[M,A]}=X \text{ and } \ex{M}=Y,\label{eq:fMXY}
\end{eqnarray}
where the optimization is carried out over a density matrix $\varrho,$ which is PPT with respect to all bipartitions.
Clearly, $f_M(X,Y)$ can be obtained via semidefinite programming.
Note the important property that $f_M(X,Y)$ is convex in $X$ and $Y,$ since the set of PPT states is a convex set.
Then, the minimum of \EQ{eq:vartheta} for a given $M$ and for PPT states is
\be\label{eq:varthetaXY}
\va{\theta} = \min_{X,Y} \frac{f_M(X,Y)-Y^2}{X^2},
\ee
which needs an optimization over two real parameters.
$\qed$

If the measured operator $M$ is known then Observation 1  provides a straightforward method to decide whether PPT states can outperform separable states.  Based on \EQS{eq:cramerrao} and \eqref{eq:Fsep}, we have to simply check whether $1/\va{\theta}\le N$ can be violated.  

We now encounter the problem of how to obtain the optimal $M$ for which we can expect a violation of the bound corresponding to separable states. Next, we will present a very efficient solution to this problem 
\cite{[{A method maximizing the QFI over noisy states, which does not use semidefinite programming, is given in }] [{.}]Macieszczak2013Quantum_arxiv}.

{\it Iterative Method.}---To find a violation of the separability bound for the quantum Fisher information with PPT states of local Hilbert space dimension $d$, we use the following iterative procedure.

\begin{enumerate}

\item  Set $j=0.$ Generate randomly a measurement operator $M.$  Set $X$ to the average of the minimum and maximum eigenvalues of the expression $i[A,M].$

\item  Compute $f_M(X,Y)$ from Eq.~(\ref{eq:fMXY}) for $Y=0$. This is a semidefinite program, which returns the optimal state $\varrho_j$. The optimal precision is
$\va{\theta}_{\varrho_j} = {\rm Tr}{(M^2\varrho_j)}/X^2$ [c.f. \EQ{eq:vartheta}].

\item Find the operator $M$  that achieves the highest $\va{\theta}_{\varrho_j}$ value for given $\varrho_j$. It is given by the symmetric logarithmic derivative~\cite{Paris2009QUANTUM}
\begin{equation}
\label{eq:SLD}
M=2i\sum_{k,l}\frac{\lambda_{k}-\lambda_{l}}{\lambda_{k}+\lambda_{l}} \vert k \rangle \langle l \vert \langle k \vert A \vert l \rangle,
\end{equation}
where $\lambda_k$ and $\ket{k}$ are now the eigenvalues and eigenvectors, respectively, of $\varrho_j.$
The quantum Fisher information of $\varrho_j$ can be obtained with $M$ as
\begin{equation}
\mathcal{F}_{\rm Q}[\varrho_j,A]=\trace(M^2\varrho_j).
\end{equation}

\item Set $X=\ex{i[M,A]}$ and $j=j+1$.

\item Repeat steps 2-4 until convergence of the objective value $\mathcal{F}_{\rm Q}[\varrho_j,A]$ is reached.

\end{enumerate}

Note that for the operator $M$ obtained in step 3 the relation $Y=\ex{M}=0$ holds. Hence, 
when at the next iteration the algorithm reaches step 2, requiring $\ex{M}=0$
means that in the worst case the same density matrix is found again as optimal. Typically, a better one is found, which implies $\mathcal{F}_{\rm Q}[\varrho_j,A]\le 1/\va{\theta}_{\varrho_{j+1}}.$ The latter inequality, together with the Cram\'er-Rao bound \eqref{eq:cramerrao}, yields $1/\va{\theta}_{\varrho_0}\le \mathcal{F}_{\rm Q}[\varrho_0,A]\le 1/\va{\theta}_{\varrho_1}\le \mathcal{F}_{\rm Q}[\varrho_1,A]\le\ldots.$ Thus, the series
$\mathcal{F}_{\rm Q}[\varrho_j,A]$ never decreases.

\begin{figure}[t]
\centerline{
\epsfxsize6.1cm \epsffile{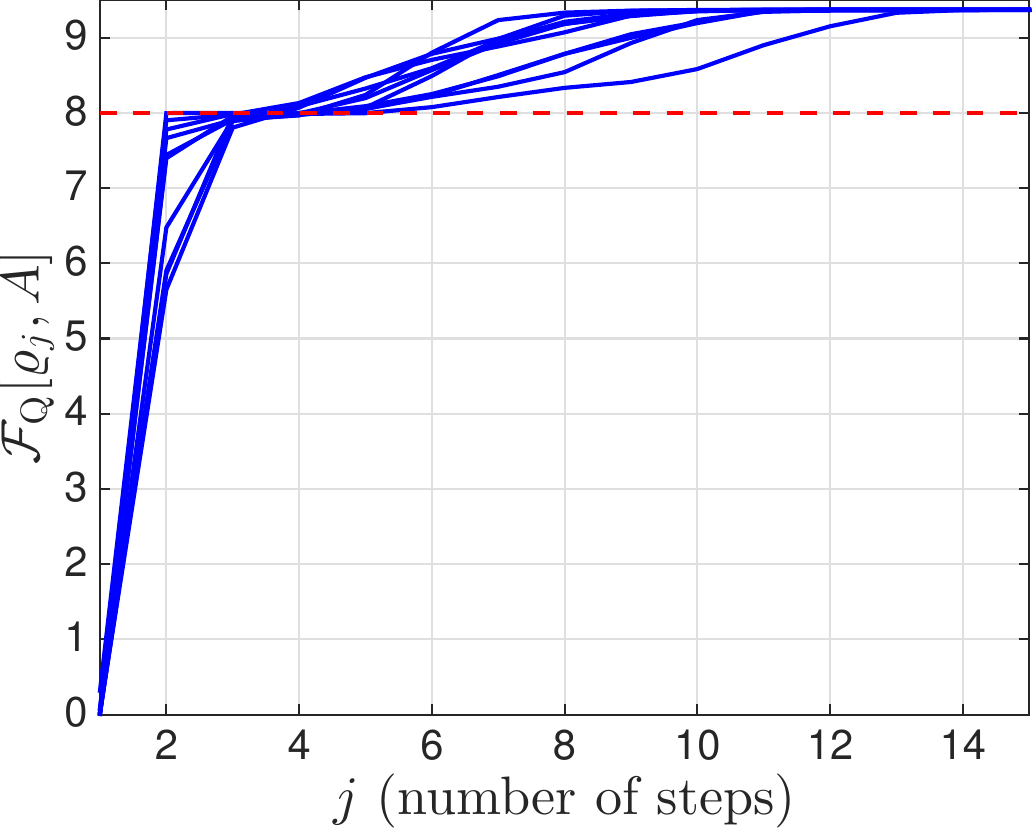}\hskip0.4cm}
\caption{Convergence to the optimal quantum Fisher information during the generation of the $4\times4$ bound entangled state, referred to also in Table~\ref{tab:FQresults_bipartite}. 
The operator $A$ is given in \EQ{eq:Doperator}.
(solid) $10$ attempts are shown. After $15$ steps, the algorithm converged to the optimal value. (dashed) The maximal value of the quantum Fisher information for separable states.}
\label{fig:convergence}
\end{figure}

{\it The rapid convergence of the algorithm.---}Our experience shows that the algorithm leads to a violation of the separable bound with $2-5$ of trials, in $10-20$ iteration steps \cite{[{For our calculations, we used MATLAB {\it version 8.4.0 (R2014b)} (The Mathworks Inc., Natick,  Massachusetts, 2014). For semidefinite programming, we used the SeDuMi, SDTP3, and YALMIP packages. See }] Vandenberghe1996Semidefinite,*Sturm2002Optimization,*TohSDPT1999,*Lofberg2004Proceedings}. We plot  the quantum Fisher information values of the density matrices obtained via the iterative algorithm for a concrete example in \FIG{fig:convergence}.

{\it Robustness of the states obtained.---}We examine how much the quantum states presented above can outperform separable states. The $12\times 12$ bipartite state, referenced in Table~\ref{tab:FQresults_bipartite}, shows a remarkably large violation of the bound corresponding to separability. The amount of violation can be characterized by the robustness of the metrological usefulness, i.e., the maximal amount of noise added for which the state performs still better than separable states. This can be obtained for white noise by direct calculation, while it can be bounded from below for PPT noise using semidefinite programming, see the Supplemental Material \CITESUPP. The robustness values obtained are given in Tables~\ref{tab:FQresults} and \ref{tab:FQresults_bipartite}. They indicate that some of our quantum states might be realized in the laboratory, since they are resistant to the level of noise present in experiments. Note that the robustness of entanglement \cite{Vidal1999Robustness} is larger or equal to the robustness based on the metrological performance.

{\it Negativity.---}So far, we have carried out an optimization for states that have a positive partial transpose for all bipartitions. The same algorithm can also be used if we relax this requirement to requiring that the smallest eigenvalue of the partial transposes is larger than $\lambda_{\rm min},$ where $\lambda_{\rm min}$ can now be negative. It is also possible to put a constraint on the negativity of the quantum state \cite{Vidal2002Computable}. In the multipartite case, we can constrain the minimum of the bipartite negativities. To this end, we use a semidefinite program. The results can be found in the Supplemental Material \CITESUPP.

{\it Metrologically useful quantum states with a local hidden variable model.}---As discussed in \REF{Czekaj2015Quantum}, it is an important question in entanglement theory whether states with another form of weak entanglement, i.e., entangled states with a local hidden variable model can also be useful for metrology. We will answer the question affirmatively.

First, we describe an example with a positive partial transpose. For that, we consider the $2\times 4$  state listed in Table~\ref{tab:FQresults}. We found that it is possible to construct numerically a local hidden variable model for the state using the algorithm of \REFS{Hirsch2016Algorithmic,Cavalcanti2016General}. In \FIG{fig:sets}, such states correspond to the set $\mathcal{P}\cap \mathcal{L}\backslash \mathcal{M},$ where ``$\cap$'' denotes the intersection of two sets.

Next we present non-PPT examples. Direct calculation shows that the two-qubit Werner state $p\ketbra{\Psi^{-}} +(1-p)\openone/4,$ defined in \REF{Werner1989Quantum}, with $\ket{\Psi^{-}}=(\ket{01}-\ket{10})/\sqrt{2}$ for $p>0.6404$ is metrologically more useful than separable states, i.e., $\mathcal{F}_{\rm Q}>2.$ We considered the dynamics given by $A=j_z\otimes \openone-\openone \otimes j_z.$ Such a state does not violate any Bell inequality for $p \le 0.68289$  using projective measurements \cite{Hirsch2017Better,Acin2006Grothendieck}.  In \FIG{fig:sets}, metrologically useful Werner states with a local hidden variable model correspond to $\mathcal{L}\backslash\mathcal{P}\backslash\mathcal{M}.$ A subset of the states above, i.e., Werner states for $p\le5/12\approx 0.4167,$ do not violate any Bell inequality, even if positive operator valued measures (POVMs) are allowed \cite{Barrett2002Nonsequential}.

{\it Conclusions.---}We showed that quantum states with a positive partial transpose  can outperform separable states in the most general metrological task of estimating a parameter in linear interferometers. A powerful iterative method was presented for finding such states. We provided examples for multipartite systems, where all the partial transposes were positive. We also presented bipartite examples. Moreover we presented PPT entangled states, as well as non-PPT entangled states, that do not violate any Bell inequality while they are still useful metrologically. 

We thank I.~Apellaniz, O.~G\"uhne, M.~Kleinmann, J.~Siewert, and G.~Vitagliano for discussions. We thank J. Ko\l odinski for drawing our attention to the reference \cite{Macieszczak2013Quantum_arxiv}. We acknowledge the support of the  EU (ERC Starting Grant 258647/GEDENTQOPT, COST Action CA15220, QuantERA CEBBEC), the Spanish Ministry of Economy, Industry and Competitiveness and the European Regional Development Fund FEDER through Grant No. FIS2015-67161-P (MINECO/FEDER, EU), the Basque Government (Grant No. IT986-16), and the National Research, Development and Innovation Office NKFIH (Grant Nos.  K124351, K111734, and KH125096).

\bibliography{pptmetro_notes,Bibliography2}

\clearpage 

\renewcommand{\thefigure}{S\arabic{figure}}
\renewcommand{\thetable}{S\arabic{table}}
\renewcommand{\theequation}{S\arabic{equation}}

\stepcounter{myfigure}
\stepcounter{mytable}
\stepcounter{myequation}
\setcounter{page}{1}
\thispagestyle{empty}


\onecolumngrid
\begin{center}
{\large \bf Supplemental Material for \\``Quantum states with a positive partial transpose are useful for metrology''}

\bigskip
G\'eza T\'oth$^{1,2,3}$ and Tam\'as V\'ertesi$^4$

\smallskip
{\it \small
$^1$Department of Theoretical Physics,
University of the Basque Country
UPV/EHU, P.O. Box 644, E-48080 Bilbao, Spain

$^2$IKERBASQUE, Basque Foundation for Science,
E-48013 Bilbao, Spain

$^3$Wigner Research Centre for Physics, Hungarian Academy of Sciences, P.O. Box 49, H-1525 Budapest, Hungary

$^4$Institute for Nuclear Research, Hungarian Academy of Sciences, P.O. Box 51, H-4001 Debrecen, Hungary}

(Dated: \today)

\medskip
\medskip

\parbox[b][1cm][t]{0.85\textwidth}{\quad
The supplemental material contains some additional results helping to characterize
the bound entangled states found numerically, as well as results of the maximization of the quantum Fisher information for a constrained negativity. We also provide a list of quantum states that are available from the electronic supplement as text files.
}

\medskip
\medskip
\medskip

\end{center}

\twocolumngrid

{\it Maximum of the quantum Fisher information for separable states.---}For any of the single qudit operators $a^{(n)}$ we have 
\bea
\va{a^{(n)}}&=& \min_{\mu}\exs{(a^{(n)}-\mu\openone)^2}\nonumber\\
 &\le & \min_{\mu} \lambda_{\max}[(a^{(n)}-\mu\openone)^2]\nonumber\\
 &=&[\lambda_{\max}(a^{(n)})-\lambda_{\min}(a^{(n)})]^2/4.\label{eq:varbound}
\eea
The first equality in \EQ{eq:varbound} is a well known identity for the variance. The inequality is based on the idea that an expectation value of an operator is never larger than its largest eigenvalue. In the second line, 
the $\mu$ leading to the minimum is $\mu=[\lambda_{\max}(a^{(n)})+\lambda_{\min}(a^{(n)})]/2.$ A state maximizing the variance, and hence saturating the inequality, is the equal superposition of the eigenstates corresponding to the minimal and maximal eigenvalues, respectively. Then, for a pure product state $\ket{\Psi}_{\rm prod}=\ket{\psi}^{(1)}\otimes\ket{\psi}^{(2)}\otimes ... \otimes\ket{\psi}^{(N)}$ we have
\bea
\mathcal{F}_{\rm Q}[\ket{\Psi}_{\rm prod},\sum_n a^{(n)}]&=&4\sum_n \va{a^{(n)}} \nonumber\\
&\le& \sum_n [\lambda_{\max}(a^{(n)})-\lambda_{\min}(a^{(n)})]^2,\nonumber\\ \label{eq:sepbound}
\eea
where the inequality can be saturated. \EQL{eq:sepbound} is valid for separable states due to the convexity of the quantum Fisher information.

{\it Comments on scaling.---}We examine the scaling of the metrological performance of PPT states with the number of particles. If we construct a tensor product of metrologically useful states, the quantum Fisher information scales as
\be
\mathcal{F}_{\rm Q}\left[\varrho^{\otimes M},\sum_{m=1}^M A^{[m]}\right]=\left(\mathcal{F}_{\rm Q}[\varrho,A]\right)^{M},
\ee
where $A^{[m]}$ acts on the $m^{\rm th}$ copy of the state. Using the four-qubit state mentioned above we obtain for $A=J_z$
\be
\mathcal{F}_{\rm Q}=1.0022 N,
\ee
where $N$ is divisible by $4.$ Hence, we have a constant factor compared to the shot-noise limit given in \EQ{eq:Fsep}. Using the state above as an initial state, one could start a numerical maximization of $\mathcal{F}_{\rm Q}$ for PPT states. It is expected that a higher level of metrological usefulness can be achieved, since we allow PPT entanglement between the four-qubit units.

{\it Description of the SDP algorithm to compute a lower bound to $p_{\rm noise}.$---}Let us consider a $d\times d$ system, and denote the maximal quantum Fisher information achievable by separable states in this system by $\mathcal{F}_{\rm Q}^{({\rm sep})}.$ Let $\varrho$ be a quantum state for which $\mathcal{F}_{\rm Q}[\varrho,A]$ is higher than $\mathcal{F}_{\rm Q}^{({\rm sep})}$. We define the robustness of the metrological usefulness of a state as follows. It is the minimal amount of separable noise that has to be mixed with $\varrho$ in order to have $\mathcal{F}_{\rm Q}\le\mathcal{F}_{\rm Q}^{({\rm sep})}$. This definition is analogous to that of the robustness of entanglement in \REF{Vidal1999Robustness}. Mathematically, we ask for the minimal amount of $p$, denoted by $p_{\rm noise}$, such that $\mathcal{F}_{\rm Q}[\varrho(p),A]\le\mathcal{F}_{\rm Q}^{({\rm sep})}$, where $\varrho(p)=(1-p)\varrho + p\varrho_{\rm sep}$, and $\varrho_{\rm sep}$ belongs to the set of separable states.

We presented lower bounds on the noise tolerance denoted by $p_{\rm noise}^{\rm LB}$ in Table~\ref{tab:FQresults_bipartite}. 
The calculation was carried out using the following SDP 
\begin{eqnarray}
p_M(X)=\min_{\sigma}&&\,p,\nonumber\\
\textrm{s.t. }&&
p = {\rm Tr}(\sigma), \sigma\ge 0, \sigma^{\rm T1}\ge 0, \nonumber \\
&& \varrho(p) = (1-p)\varrho + \sigma, \nonumber \\
&& {\rm Tr}\{\varrho(p)i[A,M]\}=X, \nonumber \\
&& \frac{1}{\va{\theta}_{\varrho(p)}}=\frac{X^2}{{\rm Tr}[M^2\varrho(p)]}\le \mathcal{F}_{\rm Q}^{({\rm sep})}. \quad \label{eq:pMX}
\end{eqnarray}
Here, $\varrho$ is the state for which we would like to obtain a bound on the robustness. The operator $M$ is obtained according to formula~(\ref{eq:SLD}). Note that the last condition can be written as ${{\rm Tr}[M^2\varrho(p)]}\ge (X^2/\mathcal{F}_{\rm Q}^{({\rm sep})})$ to make it suitable for an SDP formulation. Then, the robustness of the quantum Fisher information is computed as
\be
p_{\rm noise}^{\rm LB}=\min_{X\in[X_{\rm min},X_{\rm max}]}{p_M(X)},
\ee
where $X_{\rm min}$ and $X_{\rm max}$ are the minimum and maximum eigenvalues of the expression $i[A,M],$ respectively.

In order to see that the SDP \eqref{eq:pMX} gives indeed a lower bound we note that (i) the state $\tilde{\sigma}=\sigma/p$ approximates the set of separable states $\varrho_{\rm sep}$ from the outside. (ii) $\mathcal{F}_{\rm Q}[\varrho(p),A]\ge 1/\va{\theta}_{\varrho(p)}.$ Both (i) and (ii) potentially increase the feasible $p$ values defined by the condition appearing in the last line of the optimization~(\ref{eq:pMX}), which entails a lower bound to $p_{\rm noise}$.

\begin{figure}[t]
\centerline{
\epsfxsize4.3cm \epsffile{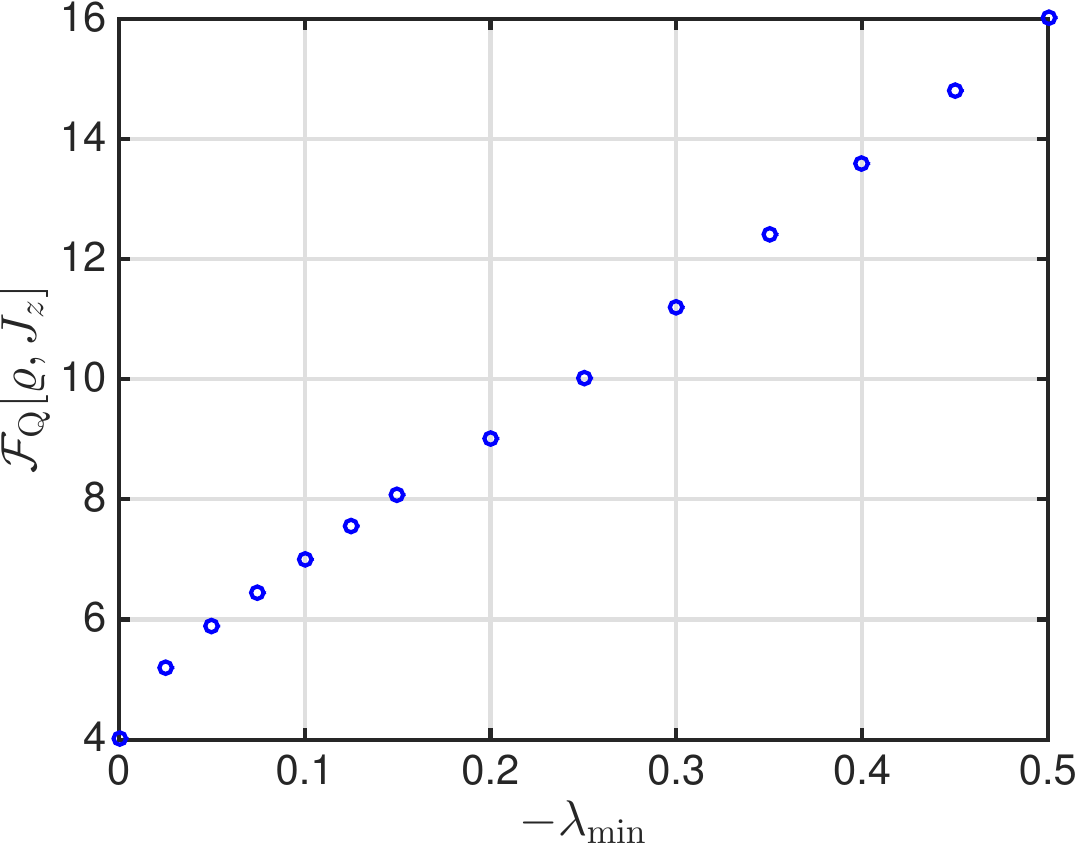}
\epsfxsize4.3cm \epsffile{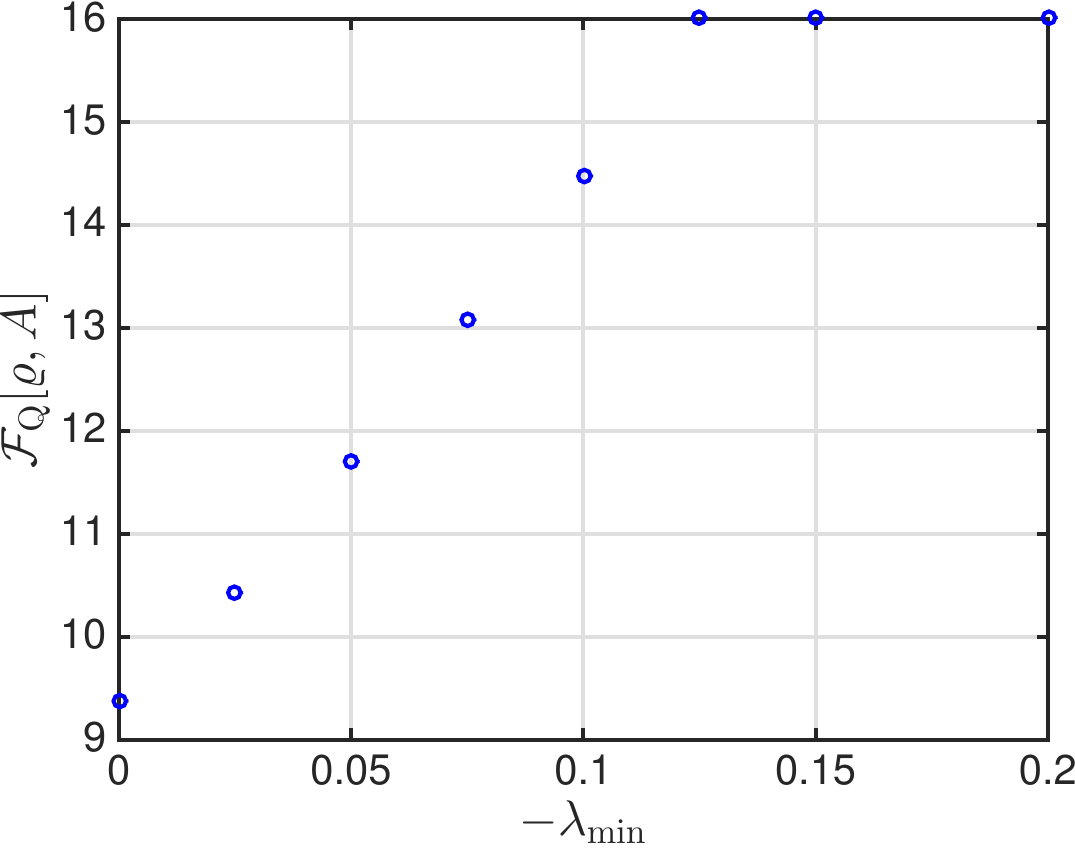}
}
\hskip0.6cm (a) \hskip4cm (b)
\caption{The maximal quantum Fisher information as the function of the smallest eigenvalue of the partial transpose.  (a) four-qubit systems and (b) $4\times4$ systems with $A$ given in \EQ{eq:Doperator}.}
\label{fig:lambdamin}
\end{figure}

\begin{figure}[t]
\centerline{
\epsfxsize6.1cm \epsffile{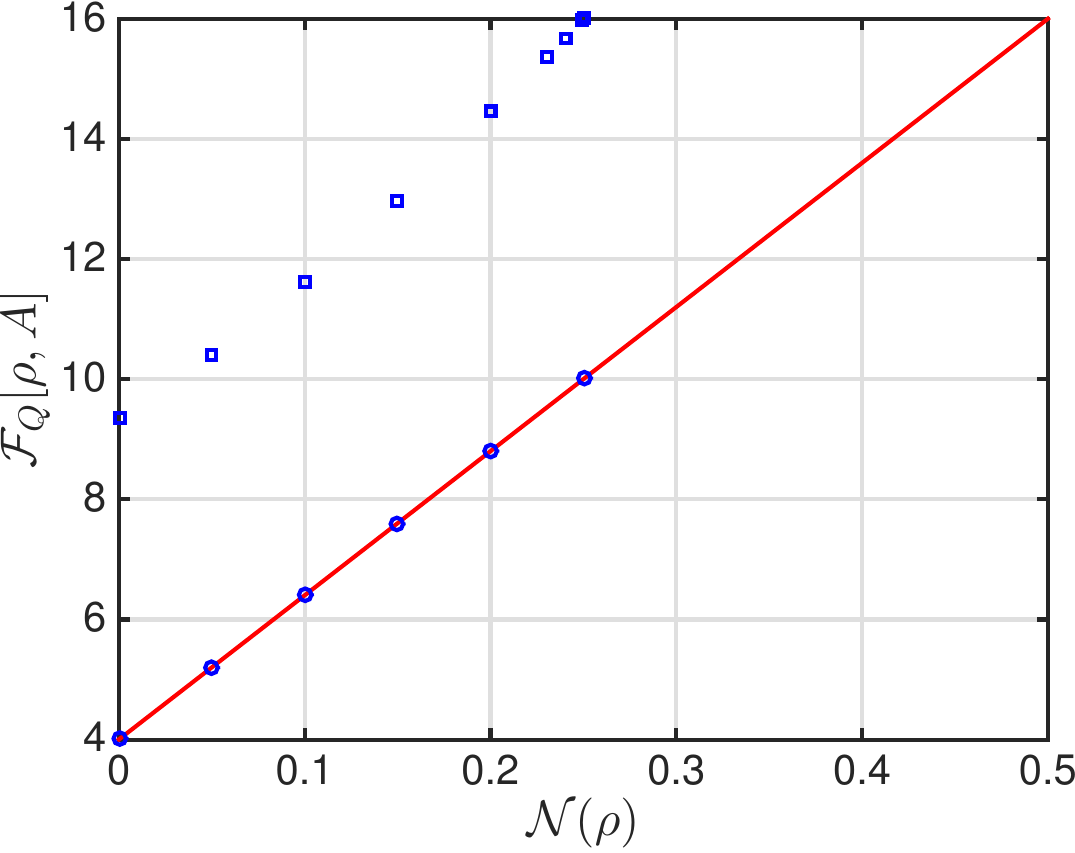}}
\caption{The maximal quantum Fisher information as the function of the smallest bipartite negativity for  (circles) four-qubit systems with $A=J_z.$  (squares) $4\times4$ systems with $A$ given in \EQ{eq:Doperator}. (solid) Line corresponding to $\mathcal{F}_{\rm Q}=24\mathcal{N}+4.$}
\label{fig:negativity}
\end{figure}

{\it Relation between the negativity and the metrological usefulness.---}We consider the constraint that the eigenvalues of $\varrho^{{\rm T}k}$ are larger than $\lambda_{\min}$ for all $k.$ We present  the maximal quantum Fisher information for various values of  $\lambda_{\rm min}$ in \FIG{fig:lambdamin}. 

We now put a constraint on the negativity of the quantum state \cite{Vidal2002Computable}. In the multipartite case, we limit the minimum of the bipartite negativities. We use the following semidefinite program \cite{Vidal2002Computable}
\begin{eqnarray}
f_M^{\mathcal N}(X,Y)=\min_{
\varrho}&&
 {\rm Tr}(M^2\varrho),\nonumber\\
\textrm{s.t. }&&
\varrho =  \varrho_+-\varrho_-,\nonumber \\
&& \varrho\ge 0,{\rm Tr}(\varrho)=1,\nonumber \\
&&{\rm Tr}(\varrho_-)=\mathcal N,\nonumber \\
&& \varrho_+^{{\rm T}k},\varrho_-^{{\rm T}k}\ge 0 \text{ for all }k,  \nonumber \\
&&  \ex{i[M,J_z]}=X \text{ and } \ex{M}=Y,\quad  \label{eq:fMXY2}
\end{eqnarray}
where the minimal bipartite negativity is not larger than ${\mathcal N}.$ We changed the original iterative algorithm by replacing $f_M(X,Y)$ defined in \EQ{eq:fMXY} by $f_M^{\mathcal N}(X,Y)$ defined in \EQ{eq:fMXY2}. The results are shown in \FIG{fig:negativity}. In the four-qubit case, the line connects our bound entangled state with the four-qubit Greenberger-Horne-Zeilinger (GHZ) state, which has a negativity of 0.5 and $\mathcal{F}_{\rm Q}[\varrho_{\rm GHZ},J_z]=16$ (see, e.g., \REF{Hyllus2012Fisher,*Toth2012Multipartite}).

\begin{table}[t]
\begin{center}
\begin{tabular}{|l|l|}
\hline
Bipartite state & Entanglement\\
\hline
\hline
$3\times3$ & $0.0003$\\
\hline
$4\times4$ & $0.0147$\\
\hline
$5\times5$ & $0.0239$\\
\hline
$6\times6$ & $0.0359$\\
\hline
$7\times7$ & $0.0785$\\
\hline
UPB $3\times3$ & $0.0652$\\
\hline
Breuer $4\times4$ & $0.1150$\\
\hline
\end{tabular}
\end{center}
\caption{Lower bound on the linear entanglement for some of the bipartite states considered in Table~\ref{tab:FQresults_bipartite}. For a comparison, the entanglement is also shown for the $3\times3$ state based on unextendible product bases (UPB) \cite{DiVincenzo2003Unextendible} and for the Breuer state with a parameter $\lambda=1/6$ \cite{Breuer2006Optimal}.
}
\label{tab:linent}
\end{table}

{\it Entanglement of the PPT entangled states.---}Next, we calculate a very good lower bound on the the entanglement measure based on the convex roof of the linear entropy of entanglement, called linear entanglement, for some of the bound entangled states presented in this paper \cite{Toth2015Evaluating}. This measure has already been used to characterize bound entangled states \cite{Sentis2016Quantitative}. The results can be seen in Table~\ref{tab:linent}. Two programs to calculate the entanglement measure are given in Ref.~\cite{pptmetro_note}. 

{\it Cluster states.}---Cluster states attracted a large attention since they can be used as a resource in measurement-based quantum computing \cite{Raussendorf2001A}. They arise naturally in Ising spin chains  and have been realized with photons and cold atoms on an optical lattice \cite{Kiesel2005Experimental,Walther2005ExperimentalOneWay, Mandel2003Controlled}. Cluster states are fully entangled pure states, hence they are not PPT with respect to any partition. They violate a Bell inequality \cite{Scarani2005Nonlocality,Guhne2005Bell,Toth2006Two-setting}. Linear cluster states of three qubits are equivalent to GHZ states under local unitaries, hence they are metrologically useful. Linear cluster states with $N\ge 4$ particles are also useful metrologically. On the other hand, for $N\ge5$ particles, ring cluster states as well as cluster states in more than one dimension are metrologically not useful  (see  Proposition~3 in \REF{Hyllus2010Not}).  In \FIG{fig:sets}, such cluster states are in  the set $\mathcal{M}\backslash\mathcal{P}\backslash\mathcal{L}.$

{\it Description of a $4\times 4$ bound entangled PPT state.---}Let us define the following six states 
$\ket{\Psi_1}=(\ket{0,1}+\ket{2,3})/\sqrt 2,$
$\ket{\Psi_2}=(\ket{1,0}+\ket{3,2})/\sqrt 2,$
$\ket{\Psi_3}=(\ket{1,1}+\ket{2,2})/\sqrt 2,$
$\ket{\Psi_4}=(\ket{0,0}-\ket{3,3})/\sqrt 2,$
and
$\ket{\Psi_5}=(1/2)(\ket{0,3}+\ket{1,2})+\ket{2,1}/\sqrt 2,
\ket{\Psi_6}=(1/2)(-\ket{0,3}+\ket{1,2})+\ket{3,0}/\sqrt 2.$
Then our $4\times 4$ state in question is a convex mixture of the following states
$\varrho_{4\times4}= p\sum_{n=1}^4\ketbra{\Psi_n}+q\sum_{n=5}^6\ketbra{\Psi_n},$
where $q=(\sqrt 2 - 1)/2$ and $p=(1 - 2q)/4$. 
The state is invariant under the partial transposition, which ensures that the state is PPT. We next show that $\rho_{4\times 4}$ is in fact a metrologically useful bound entangled state. 
We consider the operator $A=H\otimes\eins+\eins\otimes H,$ where $H={\rm diag}(1,1,-1,-1)$. 
For the $\varrho_{4\times4}$ state, $\bra{\Psi_k}A\ket{\Psi_l}=0$ for all $k,l=1,2,\ldots,6.$ Straightforward calculations show that this property implies the equality $\mathcal{F}_{\rm Q}[\varrho,A]=4(\Delta A)^2.$ 
Then, we obtain $\mathcal{F}_{\rm Q}[\varrho,A]=4(\Delta A)^2=32-16\sqrt 2\simeq9.3726$. Since for separable states $\mathcal{F}_{\rm Q}[\varrho_{\rm sep},A]\le 8$ holds, we find that the state $\varrho_{4\times4}$ is indeed bound entangled.

{\it Wigner-Yanase skew information.---}There are alternatives of the quantum Fisher information, that, apart from a constant factor, coincide with it for pure states and are convex \cite{Petz2008Quantum,Toth2013Extremal}. The Wigner-Yanase skew information $I(\varrho,A)=\trace({A^2\varrho-A\sqrt{\varrho}A\sqrt{\varrho}})$ is such a quantity \cite{Wigner1963INFORMATION}. The limit for separability for $I(\varrho,A)$ is the same as for $\mathcal{F}_{\rm Q}[\varrho,A]/4,$ since in general $I(\varrho,A)\le \mathcal{F}_{\rm Q}[\varrho,A]/4.$ We find that even for the skew information, there are PPT entangled states that violate the separable limit. For the $4\times4$ bound entangled state presented in the previous paragraph, $\ket{\Psi_k}$ for $k,l=1,2,\ldots,6$ have been used to denote the eigenvectors of the density matrix corresponding to nonzero eigenvalues. For these, as has already been mentioned, the property $\bra{\Psi_k}A\ket{\Psi_l}=0$ holds.
Straightforward algebra shows that due to this, $I(\varrho,A)=\mathcal{F}_{\rm Q}[\varrho,A]/4=9.3726/4=2.3431,$ where $A$ is given in \EQ{eq:Doperator} and $\mathcal{F}_{\rm Q}[\varrho,A]$ is shown in Table~\ref{tab:FQresults_bipartite}. The skew information signals entanglement since the bound for separability is $2.$

{\it Quantum states obtained numerically.---}The list of quantum states submitted with the supplement are given in Table~\ref{tab:states}.

\begin{table} [b!]
\begin{center}
\begin{tabular}{|l|l|}
\hline
System & File name\\
\hline
\hline
four qubits & 
\begin{tabular}{c} \verb|rho_fourqubits_r.txt| \\ \verb|rho_fourqubits_i.txt |\end{tabular}\\
\hline
three qubits & \begin{tabular}{c} \verb|rho_threequbits_r.txt| \\ \verb|rho_threequbits_i.txt |\end{tabular}\\
\hline
$2\times4$ & \begin{tabular}{c} \verb|rho_2x4_r.txt| \\ \verb|rho_2x4_i.txt |\end{tabular}\\
\hline
$3\times3$ & \verb|rho3x3.txt|\\
\hline
$4\times4$ & \verb|rho4x4.txt|\\
\hline
$5\times5$ & \verb|rho5x5.txt|\\
\hline
$6\times6$ & \verb|rho6x6.txt|\\
\hline
$7\times7$ & \verb|rho7x7.txt|\\
\hline
$8\times8$ & \verb|rho8x8.txt|\\
\hline
$9\times9$ & \verb|rho9x9.txt|\\
\hline
$10\times10$ & \verb|rho10x10.txt|\\
\hline
$11\times11$ & \verb|rho11x11.txt|\\
\hline
$12\times12$ & \verb|rho12x12.txt|\\
\hline
\end{tabular}
\end{center}
\caption{Quantum states submitted with this work, which have appeared in Tables~\ref{tab:FQresults} and \ref{tab:FQresults_bipartite}. The elements of the density matrices are given in text files.
For the first three states, the real and imaginary parts are given in two separate files, while for the rest the imaginary part is zero.
}
\label{tab:states}
\end{table}


\eject

\end{document}